\documentclass[preprint,aps]{revtex4}

\begin{document}

\title{Localization Issues for Robertson-Walker Branes
\footnote{Proceedings of "Cosmology and Elementary Particle Physics", Coral Gables Conference, December 2001, B. N. Kursunoglu, S. L. Mintz and A. Perlmutter (Eds.), American Institute of Physics, NY (2002).}}

\author{Philip D. Mannheim}
\affiliation{Department of Physics\\
University of Connecticut\\ Storrs, CT 06269 \\
{\tt electronic address: philip.mannheim@uconn.edu}}

\date{February 21, 2002}

\begin{abstract}
We discuss some of the localization issues associated with the embedding 
of Robertson-Walker type Randall-Sundrum branes in a bulk $AdS_5$. 
Specifically, we show that of the branes which are embeddable in $AdS_5$ the
geometry associated with $M_4$ and $dS_4$ branes warps away from the brane
while that associated with $AdS_4$ and $RW$ branes of any spatial 3-curvature
antiwarps away from the brane. We discuss the gravitational fluctuations
around an $M_4$ brane and analyze the specific role played by a delta
function singularity at the brane. We show how a bulk sine-Gordon scalar
field can without any fine-tuning naturally lead to localization of gravity
around an $M_4$ brane. 
\end{abstract}

\maketitle

\section{The Randall-Sundrum Set-Up}

Recently Randall and Sundrum \cite{Randall1999a,Randall1999b} showed that in the
presence of a 5-dimensional anti-de Sitter ($AdS_5$) bulk it is possible for
gravity to localize to a lower dimensional brane embedded in it, and that such
localization could be achieved even if the bulk extra dimension was infinite.
With such an $AdS_5$ bulk the probability for propagation of gravitational
signals can fall off exponentially away from the brane, with an observer on the
brane then effectively seeing only 4-dimensional rather than 5-dimensional
gravity despite the presence of the infinite extra dimension that the bulk
possesses, to thus enable us to be living in a universe with a macroscopically
sized fifth dimension. Given such an intriguing possibility it is thus necessary
to explore just how general it might be and to ascertain in what way it might
even be amenable to experimental testing. In this paper we shall therefore
explore these issues.

It is useful to begin first with a discussion of $AdS_5$ spaces themselves and to
subsequently then discuss the embedding of branes (viz. lower dimensional
surfaces) in them. $AdS_5$ is a maximally symmetric 5-space of constant negative
curvature $-b^2$. As such its Riemann tensor is given by 
\begin{equation}
R_{ABCD}=b^2(g_{AC}g_{BD}-g_{AD}g_{BC})
\label{1}
\end{equation}
(here we use $A,B=0,1,2,3,5$ to denote the five bulk coordinates $t,x,y,z,w$ with
$\mu,\nu=0,1,2,3$ denoting the ordinary 4-dimensional spacetime coordinates on
the brane), so that the 5-dimensional Weyl tensor $C_{ABCD}$ vanishes identically
while the 5-dimensional Einstein tensor is given by $G_{AB}=-6b^2g_{AB}$. Given
Eq. (\ref{1}) it is possible to construct an explicit form for the metric on the
5-space, with the most convenient one being given by a 4-dimensional Minkowski
($M_4$) sectioning of the 5-space, viz.
\begin{equation}
ds^2=e^{-2bw}(-dt^2+dx^2+dy^2+dz^2)+dw^2~~.
\label{2}
\end{equation}
(We discuss other possible sectionings below, with each such sectioning then being
associated with 4-dimensional surfaces which can in fact be embedded in $AdS_5$.)
With the fifth coordinate $w$ ranging from $-\infty$ to $\infty$, the metric of
Eq. (\ref{2}) has two interesting aspects. First, with the metric falling
away from the brane (viz. warping) in the $w \geq 0$ region, null geodesic
(viz. $dt/dw=\exp(bw)$) signals emitted at $w=\infty$ will take an infinite
amount of time to reach $w=0$, with $w=\infty$ thus being a horizon. However,
with the metric rising away from the brane (viz. anti-warping) in the $w \leq 0$
region, geodesic signals emitted at $w=-\infty$ will be able to reach $w=0$
in the finite time $t=1/b$. Consequently new information can come in from the edge
of $AdS_5$ in a finite time, thus making it impossible to unambiguously specify
the forward propagation of Cauchy data on an initial spacelike hypersurface.
$AdS_5$ spaces are thus globally non-hyperbolic.

As first suggested by Randall and Sundrum, if we could somehow get rid of the
antiwarping region while retaining only the warping one, we would then have
localization of the geometry around $w=0$. To achieve this Randall and Sundrum
therefore suggested to replace Eq. (\ref{2}) by the $w \rightarrow -w$ $Z_2$
invariant metric
\begin{equation}
ds^2=e^{-2b|w|}(-dt^2+dx^2+dy^2+dz^2)+dw^2=
e^{-2b|w|}\eta_{\mu\nu}dx^{\mu}dx^{\nu}+dw^2~~,
\label{3}
\end{equation}
a metric which thus warps for both positive and negative $w$. Operationally, Eq.
(\ref{3}) entails keeping only the $w \geq 0$ region of Eq. (\ref{2}) while
replacing the $w\leq 0$ region by a copy of the $w \geq 0$ region, to give a $Z_2$
doubling of the $w \geq 0$ region. While such a doubling then yields warping for
all $w$, we also note that the removing of the antiwarping region from
consideration thus now gives us good Cauchy propagation of initial data as well.
The Randall-Sundrum proposal thus not only achieves localization of the geometry,
it also nicely finesses the global non-hyperbolicity problem as well. Moreover,
given the modification of Eq. (\ref{3}), the Riemann tensor now no longer
obeys Eq. (\ref{1}). Rather, it instead evaluates to
\begin{equation}
R_{ABCD}=b^2(g_{AC}g_{BD}-g_{AD}g_{BC})-
2b\delta^5_A\delta^{\mu}_B\delta^5_C\delta^{\nu}_D\eta_{\mu\nu}\delta(w)
-2b\delta^{\mu}_A\delta^{5}_B\delta^{\nu}_C\delta^{5}_D\eta_{\mu\nu}\delta(w)
\label{3a}
\end{equation}
and is thus now only a pure $AdS_5$ metric in the bulk region away from $w=0$.  

In order to see what is dynamically required to yield Eq. (\ref{3}) it is
convenient to consider a slightly more general metric than it, viz.
\begin{equation}
ds^2=e^{2f(w)}(-dt^2+dx^2+dy^2+dz^2)+dw^2~~,
\label{4}
\end{equation}
a metric whose Weyl tensor still vanishes (since the metric is conformal to
flat), but whose Einstein tensor is given by
\begin{equation}
G_{00}=-G_{11}=-G_{22}=-G_{33}=3e^{2f}f^{\prime \prime}+6e^{2f}f^{\prime 2}~~,~~
G_{55}= -6f^{\prime 2}~~.
\label{5}
\end{equation}
If we consider $f(w)$ to now be a function of $|w|$, on noting that $d|w|/dw=
\theta(w) - \theta(-w)=\epsilon(w)$, $d^2|w|/dw^2=2\delta(w)$, we see (as may be
anticipated from Eq. (\ref{3a})) that all four of the
$G_{00},~G_{11},~G_{22},~G_{33}$ components of the Einstein tensor must now
contain a delta function term, while
$G_{55}$ must not. (With $dG_{55}/dw$ being a second derivative function of $w$
as is required by the Bianchi identities, $G_{55}$ itself can only contain first
derivatives of $f(w)$.) If we now impose the 5-dimensional Einstein equations,
viz.
\begin{equation}
G_{AB}=-\kappa_5^2T_{AB}~~,
\label{6}
\end{equation}
we thus see that all four of the $T_{00},~T_{11},~T_{22},~T_{33}$ components of
the energy-momentum tensor must contain a $\delta(w)$ term while $T_{55}$ must
not. We are thus led to introduce a source of energy-momentum at $w=0$, and it
is thus at $w=0$ that we must locate a lower dimensional surface or brane
(viz. one which does not contribute to $T_{55}$), a matter bearing membrane
which is thus confined to the $w=0$ region. To this end we thus set 
\begin{equation}
T_{AB}=T^{bulk}_{AB}+\delta_A^{\mu}\delta_B^{\nu}T^{brane}_{\mu \nu}\delta(w)~~,
\label{7}
\end{equation}
and find that with the introduction of bulk and brane cosmological constants 
\begin{equation}
T^{bulk}_{AB}=-\Lambda_5g_{AB} ~~,~~T^{brane}_{\mu \nu}=
-\lambda \eta_{\mu\nu}~~,
\label{8}
\end{equation}
where $\Lambda_5$ and $\lambda$ are both required to be positive, 
the metric of Eq. (\ref{3}) then emerges as the exact solution to the Einstein
equations provided only that 
\begin{equation}
6\Lambda_5+\kappa_5^2\lambda^2=0~~,
\label{9}
\end{equation}
with the bulk (viz. the $w \neq 0$ region) then being found to be the desired
$AdS_5$  with its curvature being given by
\begin{equation}
b^2=-\Lambda_5\kappa_5^2/6~~.
\label{10}
\end{equation}
As we see, in order to implement the solution we thus need a relationship  between
$\Lambda_5$ and $\lambda$, the so-called Randall-Sundrum fine-tuning
condition, a condition without which Eq. (\ref{3}) could not otherwise have
been obtained. Having now obtained our desired warping geometry, in order to
gain further insight into it we find it very convenient to consider the
embedding aspects of the problem.

\section{Embedding a brane in a bulk}

In order to discuss the embedding of our 4-dimensional universe into a
5-dimensional bulk space with some initially completely general metric $g_{AB}$,
it is particularly convenient \cite{Shiromizu2000} to base the analysis on the
purely geometric Gauss embedding formula
\begin{equation}
^{(4)}R^{\alpha}_{\phantom{\alpha} \beta \gamma \delta}=
R^A_{\phantom{A}BCD}q_A^{\phantom{A}\alpha}
q^B_{\phantom{B}\beta}q^C_{\phantom{C}\gamma}q^D_{\phantom{D}\delta}-
K^{\alpha}_{\phantom{\alpha} \gamma}K_{\beta \delta}+
K^{\alpha}_{\phantom{\alpha} \delta}K_{\beta \gamma}~~,
\label{11}
\end{equation}   
a formula which relates the 4-dimensional Riemann tensor 
$^{(4)}R^{\alpha}_{\phantom{\alpha} \beta \gamma \delta}$ 
on a general 4-dimensional surface (one not yet $Z_2$ doubled) to the
Riemann tensor $R^A_{\phantom{A}BCD}$ of a 5-dimensional bulk (one not
necessarily $AdS_5$) into which it is embedded via a term quadratic in the
extrinsic curvature $K_{\mu\nu}=q^{\alpha}_{\phantom{\alpha}\mu}
q^{\beta}_{\phantom{\beta}\nu} n_{\beta ;\alpha}$ of the 4-surface. Here
$q_{AB}=g_{AB}-n_An_B \equiv q_{\mu\nu}$ is the  metric which is induced on the
4-surface by the embedding and is thus the one with which
$^{(4)}R^{\alpha}_{\phantom{\alpha} \beta \gamma\delta}$ is calculated, while
$n^{A}$ is the embedding normal. Equation (\ref{11}) thus shows that the
4-dimensional Riemann tensor on the surface is not simply an appropriate
projection of the 5-dimensional Riemann tensor. Rather the two tensors differ by
terms which explicitly depend on the extrinsic curvature of the surface. On
introducing the bulk Weyl tensor 
\begin{eqnarray}
C_{ABCD}=
R_{ABCD}-
(g_{AC}R_{BD}-g_{AD}R_{BC}-g_{BC}R_{AD}+g_{BD}R_{AC})/3
\nonumber \\
+R^E_{\phantom{E}E}(g_{AC}g_{BD}-g_{AD}g_{BC})/12~~,
\label{12}
\end{eqnarray}   
contraction of indices in Eq. (\ref{11}) immediately allows us to relate the 4-
and 5-dimensional Einstein tensors according to 
\begin{eqnarray}
^{(4)}G_{\mu \nu}=2G_{AB}(q^{A}_{\phantom{A}\mu}q^{B}_{\phantom{B}\nu}+
n^{A}n^{B}q_{\mu\nu})/3-G^{A}_{\phantom{A}A}q_{\mu\nu}/6
\nonumber \\
-KK_{\mu \nu}+K^{\alpha}_{\phantom{\alpha}\mu}K_{\alpha\nu}+
(K^2-K_{\alpha\beta}K^{\alpha\beta})q_{\mu\nu}/2-E_{\mu\nu}
\label{13}
\end{eqnarray}   
where
\begin{equation}
E_{\mu\nu}=C^A_{\phantom{A}BCD}n_{A}n^{C}q^B_{\phantom{B}\mu}
q^D_{\phantom{D}\nu}~~.
\label{14}
\end{equation}   
The geometric content of Eq. (\ref{13}) is, first, that of the 35 components of
$C_{ABCD}$ (viz. the 35 components of the 50 component $R_{ABCD}$ which are
independent of $G_{AB}$) 10 of them can be determined once the induced metric on
the 4-surface is known; and second, that since the left hand side of Eq.
(\ref{13}) only contains derivatives with respect to the four coordinates other
than the one in the direction of the embedding normal $n^{A}$, on the right hand
side all derivative terms with respect to this fifth coordinate must mutually
cancel each other identically. Thus for instance, for the metric of the form
$ds^2=f(w)(-dt^2+d\bar{x}^2) +dw^2$ and for normal $n^{A}=(0,0,0,0,1)$, term by
term Eq. (\ref{13}) yields
\begin{eqnarray}
^{(4)}G^0_{\phantom{0} 0}=-f^{\prime\prime}/f-f^{\prime
2}/f^2+f^{\prime\prime}/f+f^{\prime 2}/4f^2
-f^{\prime 2}/f^2
\nonumber \\
+f^{\prime
2}/4f^2+2f^{\prime 2}/f^2-f^{\prime 2}/2f^2-0~~,
\label{15}
\end{eqnarray}   
i.e. $0=0$ as is to be expected since $^{(4)}G^0_{\phantom{0} 0}$ vanishes
identically in the flat 4-dimensional Minkowski space $M_4$. Finally, the
dynamical implication of Eq. (\ref{13}) is that even if $G_{AB}$ is taken to obey
the 5-dimensional Einstein equations in the 5-space, the induced 4-dimensional
$^{(4)}G_{\mu \nu}$ would not in general be expected to obey the standard
4-dimensional ones. Consequently, the dynamical structure of embedded
4-dimensional gravity is in principle different from that of non-embedded
gravity, with measurement of $^{(4)}G_{\mu\nu}$, viz. measurement purely within
the 4-dimensional world itself, then in principle enabling us to see effects
coming from higher dimensions. Equation (\ref{13}) thus provides a 4-dimensional
window on a higher dimensional world.

In order to extend the purely geometric Eq. (\ref{13}) to the Randall-Sundrum
case of interest, we note that once some energy density is placed on the $w=0$
surface, there will then be a discontinuity in the extrinsic curvature of the
surface as it is crossed from one side to the other. And for the situation in
which the Einstein equations hold in the bulk it can be shown very generally that
this discontinuity takes the form \cite{Israel1966} 
\begin{equation}
K_{\mu\nu}(w=0^{+})-K_{\mu\nu}(w=0^{-})=-\kappa^2_5[T^{brane}_{\mu\nu}-
q_{\mu\nu}(T^{brane})^{\alpha}_{\phantom{\alpha}\alpha}/3]~~.
\label{17}
\end{equation}   
As such these Israel junction conditions constitute the general relativistic
generalization of the discontinuity in a Newtonian gravitational field as a
sheet of non-relativistic matter is crossed (viz. the direction of the field is
always toward the matter distribution). While there is a discontinuity in the
extrinsic curvature it is important to note that there is no such discontinuity
in the induced metric itself so that $q_{\mu\nu}(w=0^{+})=q_{\mu\nu}(w=0^{-})$.
To implement the $Z_2$ doubling we now take the 5-space metric to be a function
of $|w|$, and with the extrinsic curvature being related to a first derivative of
the normal, $K_{\mu \nu}(w)$ then behaves as a discontinuous
$\theta(w)-\theta(-w)$ type function, so that
$K_{\mu\nu}(w=0^{-})=-K_{\mu\nu}(w=0^{+})$. In the presence of $Z_2$ doubling
we thus obtain
\begin{equation}
K_{\mu\nu}(w=0^{+})=-\kappa^2_5[T^{brane}_{\mu\nu}-
q_{\mu\nu}(T^{brane})^{\alpha}_{\phantom{\alpha}\alpha}/3]/2~~.
\label{18}
\end{equation}   
at the brane. Now since, as we noted earlier, $^{(4)}G_{\mu\nu}$ 
involves no derivatives with respect to $w$ (i.e. like $q_{\mu \nu}$ it is
continuous at the brane), even in the event that we take $g_{AB}$ to be a
function of $|w|$, it follows that $^{(4)}G_{\mu \nu}$ cannot acquire
any $\delta(w)$ term. Consequently, the right hand side of Eq. (\ref{13}) must
also contain no net $\delta(w)$ dependent term either. However, given a generic
brane matter density
\begin{equation}
T^{brane}_{\mu\nu}=-\lambda q_{\mu\nu}+\tau_{\mu\nu}
\label{19}
\end{equation}   
it follows from Eqs. (\ref{6}) and (\ref{7}) that as far as the delta
function terms are concerned, the Einstein tensor terms in Eq. (\ref{13}) make a
contribution 
\begin{equation}
2G_{AB}(q^{A}_{\phantom{A}\mu}q^{B}_{\phantom{B}\nu}+
n^{A}n^{B}q_{\mu\nu})/3-G^{A}_{\phantom{A}A}q_{\mu\nu}/6=
-\kappa^2_5[4\tau_{\alpha \beta}q^{\alpha}_{\phantom{\alpha}\mu}
q^{\beta}_{\phantom{\beta}\nu}-\tau^{\alpha}_{\phantom{\alpha}\alpha}
q_{\mu\nu}]\delta(w)/6
\label{20}
\end{equation}   
on the brane. Since the extrinsic curvature terms contain no $\delta(w)$ terms
($T^{brane}_{\mu\nu}$ is defined as the coefficient $\delta(w)$ in Eq.
(\ref{7})), it then follows that on the brane $E_{\mu\nu}$ must contain a
discontinuous delta function term of the form \cite{Mannheim2001c}
\begin{equation}
E^{disc}_{\mu\nu}=
-\kappa^2_5[4\tau_{\alpha \beta}q^{\alpha}_{\phantom{\alpha}\mu}
q^{\beta}_{\phantom{\beta}\nu}-\tau^{\alpha}_{\phantom{\alpha}\alpha}
q_{\mu\nu}]\delta(w)/6,
\label{21}
\end{equation}   
a quantity that need not vanish even if the Weyl tensor vanishes in the bulk.

With the $\delta(w)$ terms in Eq. (\ref{13}) thus taking care of each other, we
can now isolate the continuous non $\delta(w)$ terms in Eq. (\ref{13}), and on
noting that any product of any two of the components of the extrinsic
curvature is itself continuous at the brane ($[\theta(w)-\theta(-w)]^2=1$), we
find that on the brane \cite{Shiromizu2000}
\begin{equation}
^{(4)}G_{\mu \nu}=\Lambda_4q_{\mu \nu}-8\pi G_{N}\tau_{\mu\nu}
-\kappa^4_5\pi_{\mu\nu}-\bar{E}_{\mu\nu}
\label{22}
\end{equation}   
where
\begin{eqnarray}
\Lambda_4=\kappa^2_5(6\Lambda_5+\kappa^2_5\lambda^2)/12~~,~~8\pi
G_{N}=\lambda\kappa^4_5/6~~,
\nonumber \\
\pi_{\mu\nu}=-\tau_{\mu\alpha}\tau_{\nu}^{\phantom{\nu}\alpha}/4
+\tau^{\alpha}_{\phantom{\alpha}\alpha}\tau_{\mu \nu}/12
+q_{\mu\nu}\tau_{\alpha\beta}\tau^{\alpha\beta}/8
-q_{\mu\nu}(\tau^{\alpha}_{\phantom{\alpha}\alpha})^2/24~~,
\label{23}
\end{eqnarray}   
and where $\bar{E}_{\mu\nu}=[E_{\mu\nu}(w=0^{+})+E_{\mu\nu}(w=0^{-})]/2$ is the
piece of $E_{\mu\nu}$ which is continuous at the brane. As such Eq. (\ref{22}) is
the equation obeyed by the Einstein tensor on the brane, and through the presence
of the $\bar{E}_{\mu\nu}$ and $\pi_{\mu\nu}$ terms we thus see an explicit
departure from the standard 4-dimensional Einstein equations associated with a
gravitational coupling constant $8\pi G_{N}=\lambda\kappa^4_5/6$ \cite{footnote1}.
Now while Eq. (\ref{22}) is completely general and does not
require any a priori assumptions regarding the geometry in the bulk, in the event
that the bulk is taken to be $AdS_5$, the continuous piece of the Weyl tensor will
then vanish and we will be able to drop the $\bar{E}_{\mu\nu}$ term altogether,
to then yield  
\begin{equation}
^{(4)}G_{\mu \nu}=\Lambda_4q_{\mu \nu}-8\pi G_{N}\tau_{\mu\nu}
-\kappa^4_5\pi_{\mu\nu}~~,
\label{24}
\end{equation}   
with the only departure from standard gravity then being through the
presence of the term quadratic in the energy density, a term which could 
potentially be of major concern in the early universe when the energy density is
large. 

Now even if we start off with an $AdS_5$ bulk, as soon as we put some additional
matter density on the brane, that matter density will immediately set up a new
gravitational field in the bulk to not only potentially modify the bulk geometry
but to also possibly delocalize gravity as well. To avoid this we must thus only
put matter densities on the brane for which Eq. (\ref{24}) then yields a 4-metric
which is embeddable in $AdS_5$, i.e. a metric which can be associated with a
sectioning of $AdS_5$. As we will see below this precisely can occur for de
Sitter, anti de Sitter and Robertson-Walker (collectively Robertson-Walker type)
branes, viz. those highly symmetric branes of relevance to cosmology. To see how
severe a constraint the very structure of the embedding actually imposes, we
note that Eqs. (\ref{6}) and (\ref{7}) actually admit of the exact solution
\cite{Brecher2000}  
\begin{equation}
ds^2=e^{-2b|w|}[-(1-2MG/r)dt^2+dr^2/(1-2MG/r)+r^2d\Omega]+dw^2~~,
\label{25}
\end{equation} 
when the brane is taken to have a Ricci flat Schwarzschild geometry. Moreover, in
this solution every single term in Eq. (\ref{22}) vanishes identically on the
brane. However, inspection of the bulk geometry in this solution shows that the
bulk Weyl tensor does not vanish off the brane (cf. $C_{0101}=2MGe^{-2b|w|}/r^3$).
Thus even though the $C_{5\mu 5\nu}$ components of the Weyl tensor needed for
$\bar{E}_{\mu\nu}$ do vanish, its other components do not, with the bulk thus
not being $AdS_5$ in this particular case, and with the Schwarzschild metric thus
not being embeddable in $AdS_5$. In and of itself then requiring the bulk Einstein
tensor to obey $G_{AB}=\kappa_5^2 \Lambda_5g_{AB}$ is thus not sufficient to force
the bulk Weyl tensor to vanish, and thus not sufficient to ensure that the
bulk be $AdS_5$. Finally, we also note that since we cannot embed the
Schwarzschild metric in $AdS_5$, if we therefore consider a general fluctuation
due to the addition of a static mass source to a background brane whose geometry
does embed in $AdS_5$, we will find that in general the fluctuation will generate
a non-zero contribution to the Weyl tensor, to thus potentially not only modify
the geometry in the bulk but to also induce a Weyl tensor contribution on the
brane as well. As we thus see, even in the event that the background Eq.
(\ref{24}) is of the form of the standard 4-dimensional Einstein equations
(i.e. cases in which the $\pi_{\mu\nu}$ term is negligible), nonetheless the
brane fluctuations around such a background will not in fact be
standard \cite{footnote2}. (Moreover, according to Eq.
(\ref{13}) fluctuations in $K_{\mu\nu}$ are also able to contribute to the
fluctuations in the brane $^{(4)}G_{\mu \nu}$.) While we shall return to a
discussion of the structure of the associated fluctuation equation below, we
turn first to a discussion of brane backgrounds which are in fact embeddable in
$AdS_5$.

\section{Embedding of Robertson-Walker branes in A\lowercase {d}S$_5$}

For metrics which are maximally 4-symmetric in the ordinary spacetime
coordinates (viz. metrics for which $^{(4)}G_{\mu \nu}=\Lambda_4q_{\mu \nu}$) the
most general possible 5-dimensional metrics take the form 
\begin{equation}
ds^2=e^{2f(w)}[-dt^2+e^{2Ht}(dx^2+dy^2+dz^2)]+dw^2=
e^{2f(w)}q_{\mu\nu}dx^{\mu}dx^{\nu}+dw^2
\label{26}
\end{equation}
and 
\begin{equation}
ds^2=e^{2f(w)}[e^{2Hz}(-dt^2+dx^2+dy^2)+dz^2]+dw^2= 
e^{2f(w)}q_{\mu\nu}dx^{\mu}dx^{\nu}+dw^2~~,
\label{27}
\end{equation}
metrics which respectively correspond to $dS_4$ and $AdS_4$ sectionings of an
otherwise initially general 5-space. Since both of these 5-dimensional metrics
just happen to be conformal to flat for any $f(w)$, requiring their Einstein
tensors to obey 
\begin{equation}
G_{AB}=-\kappa_5^2[-\Lambda_5g_{AB}-\delta_A^{\mu}\delta_B^{\nu}\lambda q_{\mu
\nu}\delta(w)]
\label{28}
\end{equation}
will then actually force the associated bulks to be $AdS_5$, with both the $dS_4$
and $AdS_4$ branes thus being embeddable in $AdS_5$. Moreover, given the explicit
form of the brane energy-momentum tensor in Eq. (\ref{28}) the $f(w)$ coefficients
are completely determined. Thus for the $dS_4$ brane embedded in $AdS_5$ we
find that \cite{DeWolfe2000,Kim2000}
\begin{equation}
ds^2=\sinh^2(b|w|-\sigma)\sinh^{-2}\sigma[-dt^2+e^{2Ht}(dx^2+dy^2+dz^2)]+dw^2
\label{29}
\end{equation}
where $\sinh\sigma=b/H$, while for the $AdS_4$ brane embedded in $AdS_5$ we
find that \cite{DeWolfe2000}
\begin{equation}
ds^2=\cosh^2(b|w|-\sigma)\cosh^{-2}\sigma[e^{2Hz}(-dt^2+dx^2+dy^2)+dz^2]+dw^2
\label{30}
\end{equation}
where $\cosh\sigma=b/H$. Additionally, on the brane the residual
cosmological constant is given by $\Lambda_4=3H^2$ in the
$dS_4$ case and by $\Lambda_4=-3H^2$ in the $AdS_4$ case. Thus we see that
when the Randall-Sundrum fine tuning $\Lambda_4=0$ condition is not obeyed the
brane becomes either de Sitter or anti de Sitter depending on the relative
strengths of the input bulk and brane cosmological constants. 

As brane theories both of these two metrics grow exponentially as $|w|\rightarrow
\infty$ and at first sight each would appear to be of the non-localizing
anti-warping type. However the $dS_4$ brane metric has a horizon at $b|w|=
\sigma$ beyond which null geodesics can never reach the brane. Since the
function $\sinh^2(b|w|-\sigma)$ falls all the way to this horizon gravity
actually does localize \cite{Garriga2000a} in the $dS_4$ brane case. For the
$AdS_4$ brane case, while there is no such horizon ($\cosh^2(b|w|-\sigma)$ never
vanishes), nonetheless the function $\cosh^2(b|w|-\sigma)$ does initially begin
to fall before eventually turning round at $|w|=\sigma/b$ and then begin to
rise. Consequently, for small enough $H$ (viz. large $\sigma$) the horizon will
be far away from the brane and the low energy fluctuations will be quite close
to the localizing ones associated with the $H=0$ $M_4$ Minkowski brane, to thus
give an approximate or effective localization of low energy gravity on the
brane \cite{Karch2001}. In this sense then localization of gravity can be
associated with both the $dS_4$ and $AdS_4$ brane cases, though for large $H$
none of the above reasoning would apply in the $AdS_4$ case and and its
localization would be lost. (For further analysis of these two cases see also
\cite{Guth2002}.)

For maximally 3-symmetric RW branes [viz. branes with metrics which
obey Eq. (\ref{24}) with $\tau_{\mu\nu}=(\rho_m +p_m)U_{\mu}U_{\nu}
+p_mq_{\mu\nu}$] their embedding in an arbitrary 5-space yields as the most
general 5-space metric 
\begin{equation}
ds^2=-dt^2e^2(w,t)/f(w,t)+f(w,t)[dr^2/(1-kr^2)+r^2d\Omega]+dw^2~~.
\label{31}
\end{equation}
However, unlike the previous $dS_4$ and $AdS_4$ brane cases, this time
the 5-space metric is not automatically conformal to flat. In fact 10 of the
components of the Weyl tensor do not necessarily vanish (the 6
$C_{\mu\nu\mu\nu}$  with $\mu\neq\nu$ and the 4 $C_{\mu 5\mu 5}$), with all of
them being found to be proportional to 
\begin{eqnarray}
C_{0505}=
(4e^3ff^{\prime \prime}-6e^3f^{\prime 2}+4e^3fk+
6e^2fe^{\prime}f^{\prime}~~~~~~~~~~
\nonumber \\
-4e^2f^2e^{\prime\prime}
-2ef^2\ddot{f}+ef\dot{f}^2+2f^2\dot{e}\dot{f})/8ef^3~~.
\label{32}
\end{eqnarray}
Consequently this time imposing the Einstein equations is not sufficient to make
the bulk be $AdS_5$. Rather one must also require the Weyl tensor to vanish.
Explicit calculation \cite{Mannheim2001a,Mannheim2001b} then shows that this can
be done, so that maximally 3-symmetric RW metrics can indeed be
embedded in $AdS_5$. However, while it can be done, in the static
RW brane case it can only be done at a price, namely there has to
be a new fine-tuning relation between the matter fields of the theory. Since the
discussion is different in the static and non-static cases we shall discuss
the two cases separately.

For the static case first, on solving the 5-dimensional Einstein equations and on
setting the bulk Weyl tensor to zero, we find \cite{Mannheim2001a,Mannheim2001b}
that the fine-tuning condition
\begin{equation}
\kappa^2_5(\lambda+\rho_m)(-\lambda+2\rho_m+3p_m)=6\Lambda_5
\label{33}
\end{equation}   
is required of the matter fields \cite{footnote3}. On setting $\nu=(-2\kappa_5^2\Lambda_5/3)^{1/2}$
the most general solution is given in the $k=+1$ case by
\cite{Mannheim2001a,Mannheim2001b} 
\begin{equation}
f=(4/\nu^2)\sinh^2(\nu w_0/2-\nu|w|/2)~~,~~e^2/f=(4/\nu^2)\cosh^2(\nu
w_0/2-\nu|w|/2)
\label{34}
\end{equation}   
where $\coth(\nu\ w_0/2)=\kappa_5^2(\lambda+\rho_m)/3\nu$, and in the $k=-1$
case by \cite{Mannheim2001a,Mannheim2001b}
\begin{equation}
f=(4/\nu^2)\cosh^2(\nu w_0/2-\nu|w|/2)~~,~~e^2/f=(4/\nu^2)\sinh^2(\nu
w_0/2-\nu|w|/2)
\label{35}
\end{equation}   
where $\tanh(\nu w_0/2)=\kappa_5^2(\lambda+\rho_m)/3\nu$. With each of these
metrics having forms which are hybrids of both of the $dS_4$ and $AdS_4$ brane
case metrics which we presented above, and with both of them antiwarping far
from the brane, whether or not they might lead to localization of gravity is not
at all apparent. While a Karch-Randall type analysis \cite{Karch2001} has yet to
be applied to either of these two metrics, we note that localization would at
least appear possible in the $k=-1$ case since this metric has a horizon at
$|w|=w_0$, with both the $f(w)$ and $e^2(w)/f(w)$ coefficients warping all the
way to it.  

In the time dependent case the $AdS_5$ embedded solution is found to take the 
form \cite{Mannheim2001b}
\begin{eqnarray}
f(w,t)=a^2[\cosh(\nu|w|/2)-(\tau/a)\sinh(\nu|w|/2)]^2~~,
\nonumber \\
e(w,t)={1 \over [\nu^2\tau^2-\nu^2a^2-4k]^{1/2}}{df(w,t)\over dt}~~,
\label{36}
\end{eqnarray}   
where the time dependent quantities $a(t)$ and $\tau(t)$ are fixed by the 
relevant Israel junction conditions
\begin{equation}
3\nu \tau=a\kappa^2_5(\lambda+\rho_m)~~,
-3\nu[\tau\dot{a}+a\dot{\tau}]=\kappa^2_5(-2\lambda+\rho_m+3p_m)
a\dot{a}~~,
\label{37}
\end{equation}   
with Eq. (\ref{37}) itself entailing the standard covariant conservation
condition
\begin{equation}
a\dot{\rho_m}+3\dot{a}(\rho_m+p_m)=0~~.
\label{38}
\end{equation}   
With a resetting of the time according to
\begin{equation}
dt^{\prime}={2\dot{a} dt\over [\nu^2\tau^2-\nu^2a^2-4k]^{1/2}}={6da \over 
[12\Lambda_4a^2+\kappa_5^4(2\lambda\rho_m+\rho_m^2)a^2-36k]^{1/2}}~~,
\label{39}
\end{equation}   
the metric then takes the convenient form
\begin{eqnarray}
ds^2=-dt^{\prime 2}[\cosh(\nu|w|/2)-(d\tau/da)\sinh(\nu|w|/2)]^2+
\nonumber \\
a^2[\cosh(\nu|w|/2)-(\tau/a)\sinh(\nu|w|/2)]^2[dr^2/(1-kr^2)+r^2d\Omega]+dw^2~~,
\label{40}
\end{eqnarray}
with the induced metric at $w=0$ now being a standard comoving RW
one. For a perfect fluid source the Einstein tensor on the brane is given by
\begin{eqnarray}
^{(4)}G_{\mu \nu}=\kappa^2_5(6\Lambda_5+\kappa^2_5\lambda^2)q_{\mu
\nu}/12-\lambda\kappa^4_5[(\rho_m+p_m)U_{\mu}U_{\nu}+p_mq_{\mu\nu}]/6
\nonumber \\
-\kappa^4_5[2\rho_m(\rho_m+p_m)U_{\mu}U_{\nu}+\rho_m(\rho_m+2p_m)q_{\mu\nu}]/12~~,
\label{41}
\end{eqnarray}   
with a specification of an equation of state for the fluid then enabling us to
determine $a(t)$, with $\tau(t)$ then being obtainable from Eq.
(\ref{37}) \cite{footnote4}. The metric of Eq.
(\ref{40}) thus describes the most general possible embedding of a comoving RW
brane of arbitrary spatial 3-curvature $k$ in an
$AdS_5$ bulk \cite{footnote5}, and
with its dependence on $w$ being so similar to that found in the static RW case,
its localization status would appear to be comparable.  
  
\section{The Gravitational Fluctuations on an M$_4$ brane}

If a small perturbative source $S_{AB}$ is added to the background geometry
associated with Eq. (\ref{6}), this will induce a small change $\delta
g_{AB}=h_{AB}$ in the background metric $g_{AB}$ and lead to the fluctuation
equation 
\begin{equation}
\Delta G_{AB}=\delta G_{AB}+\kappa^2_5 \delta T _{AB}=-\kappa^2_5
S_{AB}~~,
\label{42}
\end{equation}   
with the associated gravitational fluctuation modes then being given as the
solutions to $\Delta G_{AB}=0$. Evaluation of Eq. (\ref{42}) for fluctuations
around an $M_4$ brane is greatly facilitated by working in the 10 condition
Randall-Sundrum gauge
\begin{equation}
h^{5A}=0~~,~~h^{\mu}_{\phantom{\mu}\mu}=0~~,~~h^{\mu
\nu}_{\phantom{\mu\nu};\nu}=0~~,
\label{43}
\end{equation}
since $\Delta G_{5A}$ is found to vanish identically in this gauge, with the
ordinary space-time components of $\Delta G_{AB}$ being found to be given by the
very compact equation \cite{Randall1999a,Randall1999b}
\begin{equation}
\Delta G_{\mu \nu}=[\partial^2_{w}-4b^2+
e^{2b|w|}\partial_{\alpha}\partial^{\alpha}
+4b\delta(w)]h_{\mu\nu}/2=-\kappa^2_5S_{\mu\nu}~~,
\label{44}
\end{equation}
an equation which is conveniently diagonal in the $\mu,\nu$ indices.
In terms of the mixed components $h^{\mu}_{\phantom{\mu}\nu}=
g^{\mu\alpha}h_{\alpha\nu}=
\exp(2b|w|)h_{\mu\nu}$ Eq. (\ref{44}) may be rewritten as 
\begin{equation}
\Delta G^{\mu}_{\phantom{\mu}\nu}=[\partial^2_{w}
-4b\epsilon(w)\partial_{w}
+e^{2b|w|}\partial_{\alpha}\partial^{\alpha}]
h^{\mu}_{\phantom{\mu}\nu}/2
=g^{-1/2}\partial_A
g^{1/2}\partial^A
h^{\mu}_{\phantom{\mu}\nu}/2=-\kappa^2_5S^{\mu}_{\phantom{\mu}\nu}~~.
\label{45}
\end{equation}
In this gauge then each mixed fluctuation component obeys the 5-dimensional
scalar Klein-Gordon equation. Moreover, for separable solutions we may simplify
Eq. (\ref{45}) by setting
$\partial_{\alpha}\partial^{\alpha}h^{\mu}_{\phantom{\mu}\nu}
=m^2h^{\mu}_{\phantom{\mu}\nu}$; and thus, when we restrict $h_{\mu\nu}$ to
depend on $|w|$, we find, on recalling that $d^2|w|/dw^2=2\delta(w)$, that
Eq. (\ref{45}) then yields two conditions that the allowed modes must satisfy,
viz.
\begin{equation}
\left[{d^2 \over d|w|^2}-4b{d \over d|w|}
+e^{2b|w|}m^2\right]\phi(|w|)=0
\label{46}
\end{equation}
and 
\begin{equation}
\delta(w){d\phi(|w|) \over d|w|}=0~~,
\label{47}
\end{equation}
where we use $\phi(|w|)$ to denote each $h^{\mu}_{\phantom{\mu}\nu}(|w|)$
component. Additionally, the allowed modes need to be properly orthonormalized.
Recalling that the covariant scalar product
\begin{equation}
(\phi_1,\phi_2)=\int(\phi^*_2\partial_{A} \phi_1 -
\phi_1\partial_{A}
\phi_2^*) n^{A}d\Sigma
\label{48}
\end{equation}
with timelike normal $n^A$ and spacelike hypersurface $d\Sigma$ provides a time
independent norm for any modes $\phi_1$ and $\phi_2$ which obey the curved space
Klein-Gordon equation, we see that Eq. (\ref{48}) is precisely the requisite
scalar product for the mixed modes $h^{\mu}_{\phantom{\mu}\nu}$, with their
finiteness thus requiring \cite{Guth2002}
\begin{equation}
\int_{-\infty}^{\infty} dw e^{-2b|w|} \phi_1^*(w)\phi_2(w) <\infty~~.
\label{49}
\end{equation}

Modes which obey all of Eqs. (\ref{46}), (\ref{47}) and (\ref{49}) are readily
found \cite{Randall1999a,Randall1999b,Garriga2000b}, with there being an
isolated massless bound state graviton with wave function 
\begin{equation}
\hat{\phi}_0(w,\bar{x},t)=Ne^{i\bar{p}\cdot\bar{x}- i|\bar{p}|t}~~,
\label{50}
\end{equation}
and normalization $N=b^{1/2}$, together with a massive continuum of modes which
begins at $m=0$ with wave functions
\begin{eqnarray}
\phi_m(w,\bar{x},t)=N(m)(m^2/b^{2})e^{2b|w|}[Y_1(m/b)J_2(me^{b|w|}/b)-
\nonumber \\
J_1(m/b)Y_2(me^{b|w|}/b)]e^{i\bar{p}\cdot
\bar{x}- i(p^2+m^2)^{1/2}t}
\label{51}
\end{eqnarray}
and normalization factor \cite{Garriga2000b,Guth2002}
\begin{equation}
N(m)={b^{2} \over 
2m^{3/2}[J_1^2(m/b)+Y^2_1(m/b)]^{1/2}}~~.
\label{52}
\end{equation}
With the fluctuation modes $h_{\mu\nu}$ being related to the mixed modes via
$h_{\mu\nu}=
\exp(-2b|w|)h^{\mu}_{\phantom{\mu}\nu}$, we thus see that for all the allowed
modes each associated wave function $h_{\mu\nu}$ falls off exponentially fast far
way from the brane, with localization of the geometry to the brane thus entailing
localization of gravity to the brane as well. Given the mode basis the retarded
propagator associated with the $h_{\mu\nu}$ modes is readily calculable
\cite{Garriga2000b,Giddings2000}, and can be written in the convenient form
\cite{Guth2002}
\begin{equation}
G(x,x^{\prime},w,w^{\prime})=
e^{-2b|w|}e^{-2b|w^{\prime}|}\sum_{m}\phi^{*}_m(w)
\phi_m(w^{\prime})
\Delta(x-x^{\prime},m)
\label{53}
\end{equation}
where $\Delta(x-x^{\prime},m)$ is the standard  4-dimensional flat Minkowski
retarded propagator for a field of mass $m$. With a static brane
source $S_{\mu\nu}=
\delta_{\mu}^0\delta_{\nu}^0M\delta^3(x)\delta(w)$ at the origin of coordinates
thus producing a fluctuation on the brane of the form
\begin{equation}
h_{00}(r,w=0)={\kappa^2_5M |\hat{\phi}_{0}(w=0)|^2\over 4\pi
r} +\kappa^2_5M\int_0^{\infty}  dm{|\phi_{m}(w=0)|^2e^{-mr} \over 4\pi r} ~~,
\label{54}
\end{equation}
we see that the massless graviton yields the conventional $1/r$ potential
on the brane with Newtonian coupling $8\pi G_N=\kappa_5^2b$. Additionally, for
large $r$ the continuum integral gets to be dominated by the small $m$ limit of
$\phi_m(w=0)$ (viz. $\phi_m(w=0) \sim m^{1/2}$), so that the continuum
integral then generates a non-leading $1/r^3$ potential \cite{Randall1999b}.
Low energy brane localized gravity is thus completely standard, with the
continuum of massive modes not affecting long distance low energy gravity on
the brane at all. 

Recalling that $b=(-\Lambda_5\kappa_5^2/6)^{1/2}$, we see that because of
the Randall-Sundrum fine-tuning condition of Eq. (\ref{9}) we may also set $b=
\lambda\kappa_5^2/6$. We thus find that the effective 4-dimensional Newton
constant defined by the propagator, viz. $8\pi G_N=\lambda\kappa_5^4/6$, is
precisely that obtained in Eq. (\ref{22}) via the embedding procedure. Now while
this is certainly a very desirable result since it confirms the consistency of
two different ways of defining $G_N$, the result is still somewhat puzzling since
though the  Eq. (\ref{22}) background reduces to $^{(4)}G_{\mu \nu}=0$ in
the $M_4$ brane case, nonetheless, it is not true that fluctuations around it
will obey $\Delta^{(4)}G_{\mu \nu}=-8\pi G_{N}\delta\tau_{\mu\nu}$ when a weak
source $S_{\mu\nu}=\delta\tau_{\mu\nu}\delta(w)=
\delta_{\mu}^0\delta_{\nu}^0M\delta^3(x)\delta(w)$
is introduced at $w=0$, since, as we noted earlier, the introduction of a mass
source on the brane potentially leads to changes in both 
$C_{ABCD}$ and $K_{\mu\nu}$. On denoting the net effect of such potential
changes by $\delta F_{\mu\nu}$, the lowest order brane fluctuations thus have to
generically obey the modified 
\begin{equation}
\Delta^{(4)}G_{\mu \nu}=-8\pi G_{N}\delta\tau_{\mu\nu}
-\delta F_{\mu\nu}
\label{55}
\end{equation}   
instead. Since Eq. (\ref{55}) is not a standard 4-dimensional Einstein
fluctuation equation, it is not immediately clear with what strength the
massless graviton then does couple, and we thus have to reconcile
Eqs. (\ref{22}), (\ref{54}) and (\ref{55}). In order to explicitly do this we
have found it very instructive to monitor the $\delta(w)$ contributions to the
fluctuation equation.  

Since $^{(4)}G_{\mu \nu}$ is associated with the induced metric on the
brane, and since it transforms as a rank two tensor with respect to the
background geometry, we can calculate the change $\Delta^{(4)}G_{\mu \nu}$ 
due to the change $\delta q_{\mu\nu}=h_{\mu\nu}$ in the induced metric using 
standard tensor calculus techniques. In the $h^{\mu}_{\phantom{\mu}\mu}=0$,
$h^{\mu\nu}_{\phantom{\mu\nu};\nu}=0$ gauge of interest explicit calculation
then shows that $\Delta^{(4)}G_{\mu\nu}=
\partial_{\alpha}\partial^{\alpha}h_{\mu\nu}/2$, so that Eq. (\ref{44}) may be
rewritten as
\begin{equation}
\Delta G_{\mu \nu}=[\partial^2_{w}-4b^2
+4b\delta(w)]h_{\mu\nu}/2+e^{2b|w|}\Delta^{(4)}G_{\mu\nu}
=-\kappa^2_5\delta\tau_{\mu\nu}\delta(w)~~.
\label{56}
\end{equation}
On Taylor expanding $h_{\mu\nu}(|w|)=a^0_{\mu\nu}+a^1_{\mu\nu}|w|
+a^2_{\mu\nu}|w|^2/2+...$, Eq. (\ref{56}) entails that
\begin{equation}
a^2_{\mu\nu}/2-2b^2a^0_{\mu\nu}+\Delta^{(4)}G_{\mu\nu}=0~~,~~
(a^1_{\mu\nu}+2ba^0_{\mu\nu})\delta(w)=
-\kappa^2_5\delta_{\mu}^0\delta_{\nu}^0M\delta^3(x)\delta(w)~~, 
\label{57}
\end{equation}
so that even while $S_{\mu\nu}$ contains a $\delta(w)$ term,
the equation involving $\Delta^{(4)}G_{\mu\nu}$ does not since
$\Delta^{(4)}G_{\mu\nu}$ itself possesses no $\delta(w)$ term. However, on
substituting for
$a^0_{\mu\nu}$ in the static case of interest we obtain 
\begin{equation}
\Delta^{(4)}G_{00}=\nabla^2h_{00}(w=0)/2=\nabla^2a^0_{00}/2=
-\kappa_5^2bM\delta^3(x)-ba^1_{00}-a^2_{00}/2~~, 
\label{58}
\end{equation}
which we  recognize as being of the form of Eq. (\ref{55}) with
$\kappa_5^2b=8\pi G_N$ and $\delta F_{00}=ba^1_{00}+a^2_{00}/2$.
For the massless graviton exchange contribution where 
$h_{00}(r,w)=\exp(-2b|w|)\kappa_5^2bM/4\pi r$, the Taylor series expansion
coefficients explicitly evaluate to 
\begin{equation}
a^0_{00}=\kappa_5^2bM/4\pi
r~,~a^1_{00}=-\kappa_5^2M[b^2/2\pi r
+\delta^3(x)]~,~a^2_{00}=\kappa_5^2bM[b^2/\pi r+
\delta^3(x)]~~, 
\label{59}
\end{equation}
so that $\delta F_{00}$ takes the value $-\kappa_5^2bM\delta^3(x)/2$ and is
thus explicitly non-zero. Thus finally, on inserting Eq. (\ref{59}) into Eq.
(\ref{58}) we obtain none other than
\begin{equation}
\nabla^2(\kappa_5^2bM/8\pi r)=-\kappa_5^2bM\delta^3(x)/2
\label{60}
\end{equation}
just as desired of massless graviton exchange on the brane. We thus conclude
that even though the fluctuations on the brane obey the non-standard Eq.
(\ref{55}), nonetheless, through a delicate interplay, the resulting
fluctuations turn out to still be completely canonical. Having now explored the
structure of the Randall-Sundrum set-up, we now briefly discuss how such a
set-up could be achieved dynamically; and shall thus explore the dynamics
associated with the coupling of gravity to a bulk sine-Gordon scalar field  (a
model also considered in \cite{Gremm2000}), and show
\cite{Davidson2000} how it naturally leads to Randall-Sundrum localization of
gravity without any need for fine-tuning. 
  
\section{Dynamical Localization of Gravity}
For a scalar field with potential
$V(\phi)=A^2\beta^2/8-(A^2\beta^2/8)(1+\kappa^5_2A^2/3)sin^2(2\phi/A)$
coupled to the metric of Eq. (\ref{4}) with $T_{00}=e^{2f(w)}[\phi^{\prime
2}/2+V(\phi)$], $T_{55}=\phi^{\prime 2}/2-V(\phi)$, there is an exact solution
to the 5-dimensional Einstein equations, viz. 
\begin{equation}
tan(\phi/A)=tanh(\beta w/2)~~,~~e^{f(w)}=[cosh(\beta w)]^{-A^2\kappa^2_5/12}~~.
\label{61}
\end{equation}
Here $e^{f(w)}$ peaks at $w=0$ while warping away from it, with the solution thus
representing a thick domain wall supported by a soliton. Moreover, without
assuming any input $Z_2$ symmetry, in the solution the output domain
wall nonetheless has acquired one from the underlying symmetry
structure which solitons intrinsically possess. Given the solution, if we now
take the limit $A\rightarrow 0$, $\beta\rightarrow \infty$ with $A^2\beta$ held
fixed, we find that \cite{Davidson2000}
\begin{equation}
e^{f(w)}\rightarrow e^{-(-\Lambda_5\kappa^2_5/6)^{1/2}|w|}
\label{62}
\end{equation}
which is precisely of the Randall-Sundrum form. Here 
$\Lambda_5=-\kappa^2_5A^4\beta^2/24$ is the minimum value of $V(\phi)$. In
this same limit we find that the scalar field energy density $T_{00}$ develops a 
$\lambda \delta (w)$ component where $\lambda= A^2\beta/2$, and thus on
comparing terms we naturally recover \cite{Davidson2000} the Randall-Sundrum
$6\Lambda_5+\kappa_5^2\lambda^2=0$ condition without fine-tuning.

The author wishes to thank to Drs. A. Davidson, A. H. Guth,  D. I. 
Kaiser and A. Nayeri for many helpful discussions. This work
has been supported in  part by the Department of Energy under grant No.
DE-FG02-92ER40716.00.


\begin{thebibliography}{99}
\bibitem{footnote3} In passing we note that in the
presence of  this Eq. (\ref{33}) the Einstein tensor on the brane is then given
by $^{(4)}G_{\mu\nu}=-8\pi G_{N}[(\rho_m+p_m)U_{\mu}U_{\nu}+p_mq_{\mu \nu} 
-(\rho_m+3p_m)q_{\mu\nu}/2] +O(\kappa^4_5\rho_m^2)$, with the leading order
source acting just like a perfect fluid with energy density $\rho =
3(\rho_m+p_m)/2$ and pressure $p=-(\rho_m+p_m)/2=-\rho/3$, i.e. acting just like
negative pressure quintessence. With negative brane pressure thus potentially
being able to arise due to the embedding into the bulk (the bulk stresses
maintain the negative pressure on the brane), there may thus be no need to
actually introduce any explicit 4-dimensional fluid with intrinsically negative
pressure into cosmology at all.
\bibitem{footnote4} We note that the time-time component of Eq.
(\ref{41}) takes the form $-3(\dot{a}^2+k)/a^2
=-\Lambda_4-\lambda\kappa_5^4\rho_m/6 - \kappa_5^4\rho_m^2/12$, a form we
immediately recognize as Eq. (\ref{39}); and in passing we also note that even
though Eq. (\ref{39}) is not necessarily always integrable in terms of named
functions, in the special quintessence case where $p_m=-\rho_m/3$, i.e. 
$\rho_m= B/a^2$, Eq. (\ref{39}) actually admits of an exact solution, viz.
$a^2(t^{\prime})=A+C\cosh(\gamma t^{\prime})$ where
$\gamma=(4\Lambda_4/3)^{1/2}$,
$A=(2k\lambda-B\nu^2)/\lambda\gamma^2-B/\lambda$, and
$C=(B\nu^2-2k\lambda)^2/\lambda^2\gamma^4
-B(4k\lambda-B\nu^2)/\lambda^2\gamma^2$. Since the standard purely 4-dimensional
cosmology $^{(4)}G_{\mu \nu}=-\kappa^2_4(-\lambda
q_{\mu\nu}+\tau_{\mu\nu})$ would yield
$dt^{\prime}=da/[\kappa_4^2(\lambda+\rho_m)a^2/3-k]^{1/2}$ for
the very same 4-dimensional sources, we see that there is an intrinsic difference
between standard and brane embedded cosmology.
\bibitem{footnote5} In complete analog to the situation found with regard 
to the $M_4$ and $dS_4$ branes, as soon as we take the brane metric to be
time dependent we are immediately released from fine tuning constraints.
\end{thebibliography}
\end{document}